\documentclass[final,5p,times,twocolumn]{elsarticle}

\usepackage{graphicx}
\usepackage{tabularx}
\usepackage{lineno}

\usepackage{amssymb}

\journal{Nuclear Instruments and Methods A}

\begin{document}
\linenumbers

\begin{frontmatter}



\title{Development of Edgeless Silicon Pixel Sensors on p-type substrate for the ATLAS High-Luminosity Upgrade}

\author[A,B]{G. Calderini}
\author[C]{A. Bagolini} 
\author[A]{M. Bomben}
\author[C]{M. Boscardin}
\author[D]{L. Bosisio}
\author[A]{J. Chauveau}
\author[C]{G. Giacomini}
\author[E]{A. La Rosa}
\author[A]{G. Marchiori}
\author[C]{N. Zorzi} 

\address[A]{Laboratoire de Physique Nucl\'eaire et des Hautes Energies (LPNHE) Paris, France}
\address[B]{Dipartimento di Fisica E. Fermi, Universit\'a di Pisa, Pisa, Italy}
\address[C]{Fondazione Bruno Kessler, Centro per i Materiali e i Microsistemi (FBK-CMM) Povo di Trento (TN), Italy}
\address[D]{Universit\`a degli studi di Trieste and INFN-Trieste, Italy}
\address[E]{Section de Physique (DPNC), Universit\`e de Geneve, Geneve, Switzerland}

\begin{abstract}
In view of the LHC upgrade for the High Luminosity Phase (HL-LHC), the ATLAS experiment is planning to replace the Inner Detector with an all-Silicon system.
The n-in-p bulk technology represents a valid solution for the modules of most of the layers, given the significant radiation hardness of this option and the reduced cost. 
The large area necessary to instrument the outer layers will demand to tile the sensors, a solution for which the inefficient region at the border of each sensor needs to be reduced to the minimum size. This paper reports on a joint R\&D project by the ATLAS LPNHE Paris group and FBK Trento on a novel n-in-p edgeless planar pixel design, based on the deep-trench process available at FTK.   

\end{abstract}

\begin{keyword}
TCAD simulations, Planar Silicon radiation detectors, fabrication technology, tracking detectors



\end{keyword}

\end{frontmatter}

\linenumbers

\section{Introduction}
To extend the physics reach of the LHC, accelerator upgrades are planned during the next ten years which will finally increase the integrated luminosity to more than 3000 fb$^{-1}$ and the pile-up per bunch-crossing by a factor 5 to 10. The price of this luminosity improvement will be a severe increase in occupancy and un-precedented radiation damage conditions which will require to upgrade the detectors and in particular their tracking systems.    
The ATLAS experiment has already defined a roadmap, based on two main steps. During the first phase, in the Long Shutdown 1 in 2013-2014 an additional pixel layer (the Insertable B-layer, IBL) \cite{TDR} will be added between the beam-pipe and the present layer-1 of the pixel detector. Later, around 2020, when the high-luminosity phase of LHC (HL-LHC) \cite{hl-lhc} will start, the whole replacement of the Inner Detector with an all-silicon system is foreseen. Simulations indicate that an integrated  fluence of about $10^{16} n_{eq}/cm^2$ is expected for the innermost pixel layer at the end of the HL-LHC phase, dropping at the level of $10^{15} n_{eq}/cm^2$  for most of the middle layers. A big R\&D effort is underway in the LHC physics community, and in particular in the framework of the ATLAS experiment, to design and produce sensors matching the HL-LHC requirements in terms of radiation hardness, material budget and segmentation at an affordable cost. In addition to planar sensors (for an updated status report see for instance \cite{pps}) other options are evaluated: Silicon 3D \cite{3d}, diamond \cite{diamond} and HV-CMOS \cite{hv-cmos} sensors. 

The future material budget limitations will demand the tiling of modules in the mid- and outer-layers and this imposes a high geometrical acceptance for the new sensors. The inefficient region along the border should be reduced to less than 2.5\% of the total surface \cite{TDR}. 
The need to reduce as much as possible the size of the dead region at the border of the sensors has driven the planar pixel community to use intensely the device simulation tools available and optimise the sensor layout. Good results have been achieved by reducing the guard ring region, which represents a low-efficiency portion of the sensor due to the lower electrical field and the distance from the first row of pixels. This has already been done the ATLAS Insertable B-Layer sensors, since preliminary simulations (see for instance \cite{hiroshima_2009}) had indicated that the number of guard rings could be reduced; the n-in-n nature of the sensors, in which pixels and guard rings are on opposite faces, has also allowed to push the first row of pixels inside the guard-ring area. 

The new sensors used in the HL-LHC upgrade will be possibly based on n-in-p technology, much cheaper with respect to the n-in-n given the smaller number of masks necessary in the production. Recent available industrial processes to achieve n-in-p sensors with virtually no border inefficiency are presented.   

\section{Active edge sensors}
In conventional pixel or strip sensors, a portion of the surface close to the edges is not instrumented and is reserved for guard-ring implants. This solution is necessary to modulate the potential at the border of the sensor, in order to prevent the electrical field from reaching the side, where the defects caused by the cut-line would act as current generators. At the same time the charge collection in this region is very limited, due to the distance from the first row of pixels and the small electrical field. In the sensors of the ATLAS pixel production, the un-instrumented region running along the edge of the sensor is 1.1 mm wide, while for the IBL it has been reduced to 
$450~\mu m$. An alternative way to avoid the generation of edge current would be to enforce an equipotential zone across the sidewalls of the sensor. In n-in-p sensors this volume should be set at the same potential as the backside, virtually extending the back to the edge in a single region. The absence of electrical field can avoid the generation of surface current even in the absence of guard-rings. A few sensor providers have developed technological processes which can achieve this result in different ways. A possibility, chosen by Fondazione Bruno Kessler (FBK) \cite{fbk, fbk-active-edge} in Trento, Italy and by VTT \cite{vtt, mpp-active-edge} Finland, is to use a deep vertical trench etched along the device periphery throughout the entire wafer thickness (this requires a support wafer to hold the sensors during the process). The edge of the trench is heavily doped via diffusion (FBK) or direct implantation (VTT) thus virtually extending the ohmic back-contact to the sensor side. A different technique is used by the SCIPP ATLAS group \cite{scipp}, in Collaboration with NRL \cite{nrl}, by using a Scribe-Cleave-Passivate (SCP) sequence to obtain a fixed interface charge in the sidewall allowing to control the potential in the edge (see for instance \cite{vitaliy}). A small number of guard rings is usually kept to improve the behaviour of the sensors at high voltage after high irradiation. 
        
\section{The active edge sensor fabrication at FBK}  
\begin{figure}[h!] 
\centering 
\includegraphics[width=0.8\columnwidth,keepaspectratio]{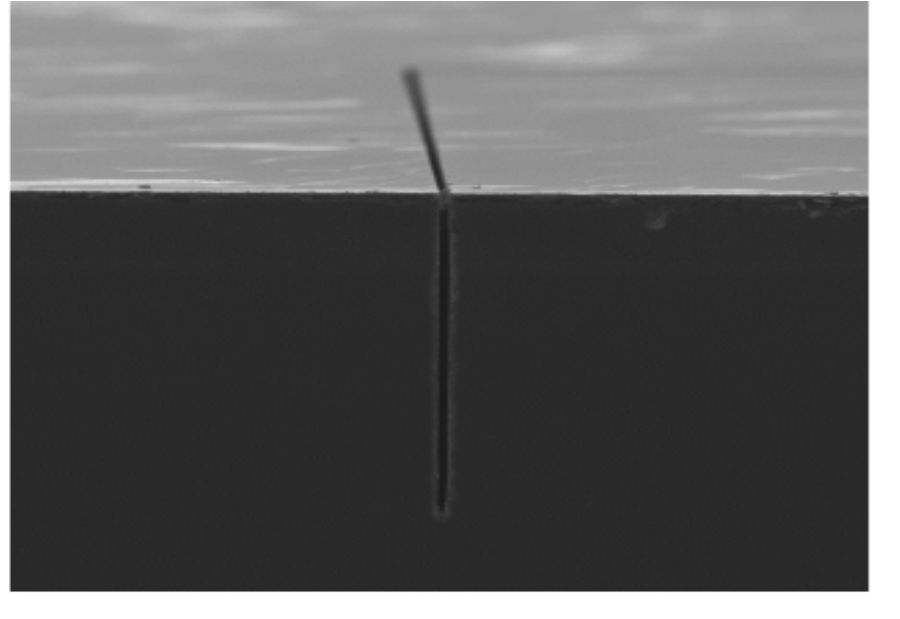}
\caption{Detail of the SEM picture of a test trench produced by DRIE.}
\label{fig:fbk-active-edge}
\end{figure}

\begin{figure}[h!] 
\centering 
\includegraphics[width=0.9\columnwidth,keepaspectratio]{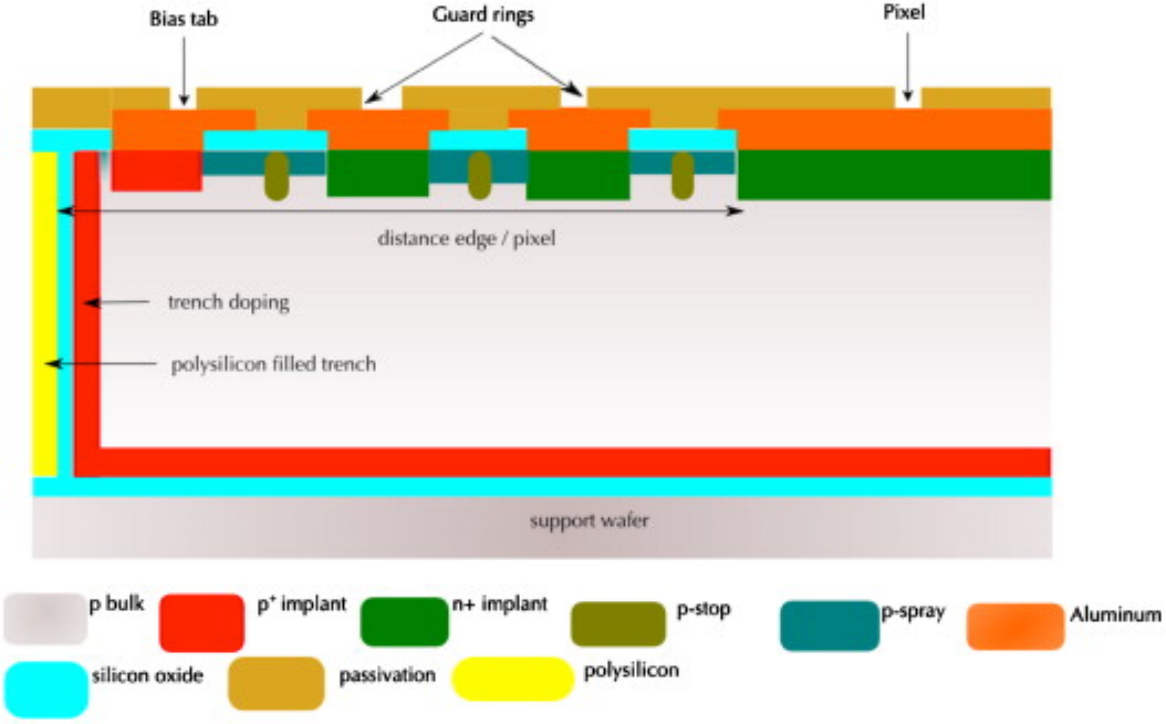}
\caption{Schematic view of the sensor structure in proximity of the edge. The bias tab, the guard rings (when present) and the first pixel are sketched.}
\label{fig:edgeless-sketch}
\end{figure}

The sensors are fabricated on 4-inch high-resistivity p-type FZ, $<100>$ oriented $200~\mu m$-thick wafers. A deep trench running along the border of each sensor is excavated by the use of Deep Reactive Ion Etching (DRIE) all the way through the wafer thickness. A $500~\mu m$ Silicon substrate is used to provide mechanical support to the devices during the process making the backside inaccessible. For this reason the back $p^+$ implantation is made as first step at FBK before the wafers are shipped to Sintef \cite{sintef} for wafer-bonding. All the subsequent process steps are made in the FBK cleanroom. A dedicated ohmic contact ("bias tab") reaching the backside is present on the front pixel face to allow to bias the detector for tests purpose even before the removal of the support wafer. 
The trench etching, the subsequent boron-doping of its walls in a diffusion furnace and the final poly-silicon filling represent critical steps of the process. FBK technology can obtain very uniform, well defined and narrow trenches (see Fig. \ref{fig:fbk-active-edge}), with a typical width of $5~\mu m$ over a depth of more than $200~\mu m$.      
Any residual of air inside the poly-silicon filling could severely damage the devices during the high-temperature subsequent steps of the fabrication. The sketch of the final device is shown if Fig. \ref{fig:edgeless-sketch}. The n-type pixels are electrically isolated by both homogeneous "p-spray" and patterned "p-stop" implants with various combinations of presence/absence and doses in different process splitting. The different options will allow to better study the junction isolation and breakdown voltage after irradiation of the devices. For test purpose, a temporary metal layer is present over the passivation oxide and patterned in such a way to connect together all the pixels of each row. In such a way the sensors can be easily biased for test before the bump-bonding to the readout electronics. Finally the metal will be removed by wet etching. 
      
\section{The wafer layout}
\begin{figure}[h!] 
\centering 
\includegraphics[width=0.7\columnwidth,keepaspectratio]{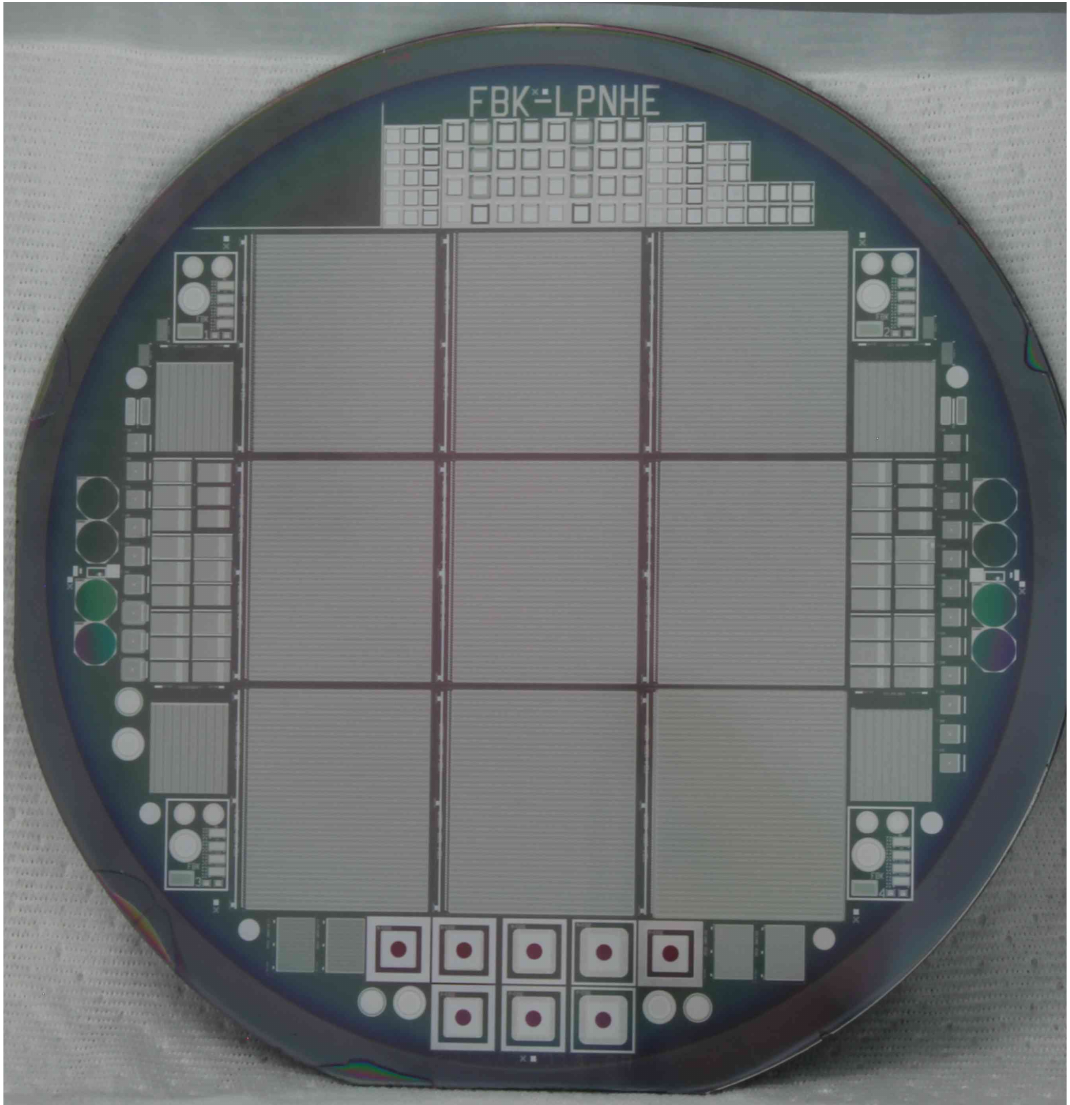}
\caption{Picture of the wafer layout. The nine large FE-I4-compatible sensors are visible at the center of the wafer. Other smaller and baby sensors are accomodated in the surrounding regions, together with a large number of test structures.}
\label{fig:layout}
\end{figure}

The FE-I4 \cite{fe-i4} compatible pixels sensors consist of an array of 336 x 80 pixels with a pitch of $50~\mu m \times 250~\mu m$ for an overall sensitive area of 16.8 mm $\times$ 20.0 mm. We placed nine FE-I4 compatible pixel sensors in the centre of the 4-inch wafer (see Fig. \ref{fig:layout}). 

They differ for pixel-to-trench distance (100, 200, 300 and 400 $\mu m$) and the number of guard rings surrounding the pixel area (0, 1, 2, 3, 5 and 10). Smaller FE-I3 compatible sensors are also present, together with baby-sensors, featuring a smaller number of pixels but preserving the same characteristics of the main sensors.  
Many test structures are also present which will be used for radiation damage studies and to validate the models used in the simulation of the devices. 

\section{The TCAD simulation}
The full sensor production has been planned on the basis of dedicated device numerical simulations with Silvaco \cite{silvaco} software. Several different designs have been evaluated with 2D simulations to study the effect of the number of guard rings and the distance of the first pixel from the edge for non irradiated and irradiated sensors up to a fluence of $\phi = 1 \times 10^{15} n_{eq}/cm^2$, the expected dose for the middle-layers of the new tracker at the end of the HL-LHC phase. The electrical field inside the device has been simulated for each configuration and the behaviour of the leakage current and breakdown voltage has been studied, together with the expected charge collection efficiency. The doped regions ($n^+$ for the pixels and the guard-rings and $p^+$ for the backside, the p-stop, p-spray, bias tab and trench walls) have been modeled with profile parameters and peak concentrations reported in Tab \ref{tab:parameters}.

\begin{table}[h!]
\caption{Parameters used in the simulation to model the detectors. A (D) stand for accceptor (donor) impurities.}
\centering
\footnotesize
\begin{tabularx}{\columnwidth}{llllll}
\hline
Doped & Impurity & Profile & Peak Value & Ref. value & Roloff \\
region &  & &(cm$^{-3})$ &(cm$^{-3})$ & $\mu m$ \\
\hline
Pixels / GRs & D & Gaussian & $2\times10^{19}$ & $10^{16}$ & 1.0 \\
Back & A & Gaussian & $2\times10^{19}$ & $10^{16}$ & 1.0 \\
Trench & A & erf & $2\times10^{19}$ & $10^{12}$ & 2.0 \\
Bias tab & A & Gaussian & $2\times10^{19}$ & $2\times10^{16}$ & 0.5 \\
P-spray & A & Gaussian & $5\times10^{16}$ & $7\times10^{15}$ & 0.5 \\
P-stop & A & Gaussian & $5\times10^{17}$ & $7\times10^{16}$ & 0.5 \\
\hline
\end{tabularx}
\label{tab:parameters}
\end{table}      

Oxide fixed charge density has been set to $N_f=10^{11}$cm$^{-2}$ for un-irradiated devices, increasing to $N_f=3\times10^{12}$cm$^{-2}$ for the irradiated case.  
To describe the radiation damage, the Pennicard effective model based on three deep levels in the forbidden gap \cite{pennicard} has been implemented in the software. 
Radiation-induced interface traps at the Silicon-Oxide interface are also modeled in the simulation as described in \cite{schwandt}. The model was validated with the use of irradiated n-in-p test structures from previous productions.

\subsection{Simulation of electrical properties and comparison with measurements}
\begin{figure}[h!] 
\centering 
\includegraphics[width=0.8\columnwidth,keepaspectratio]{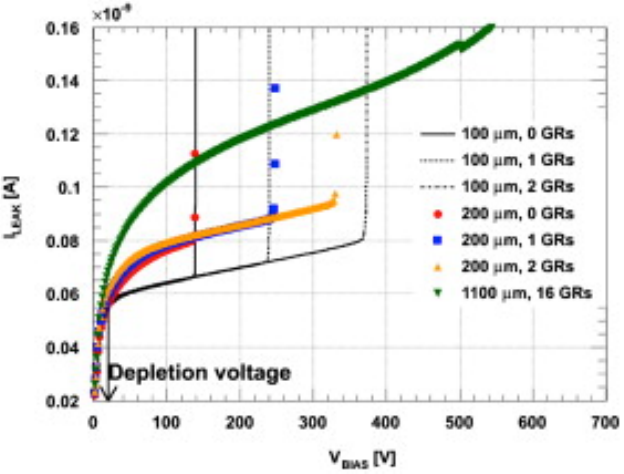}
\caption{Simulated IV behaviour of the pixels of the first row for different sensor designs before irradiation. The calculated depletion voltage is indicated by the arrow.}
\label{fig:IV-noirrad}
\end{figure}

\begin{figure}[h!] 
\centering 
\includegraphics[width=0.8\columnwidth,keepaspectratio]{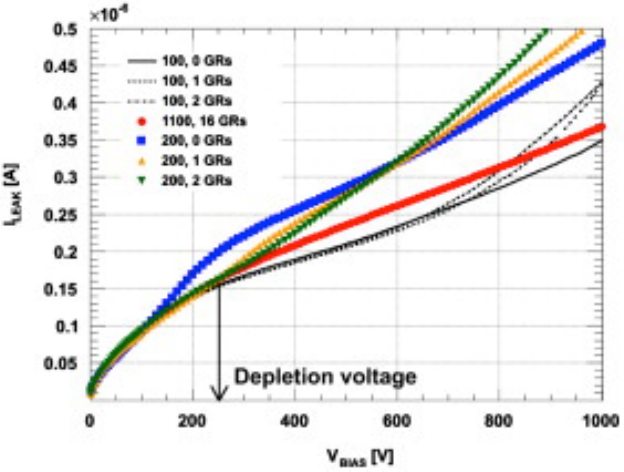}
\caption{Simulated IV behaviour of the pixels of the first row for different sensor designs after a simulated fluence of $10^{15} n_{eq}/cm^2$ based on the radiation damage models described in the text. The calculated depletion voltage is indicated by the arrow.}
\label{fig:IV-irrad}
\end{figure}

Simulated behaviour of the leakage current and breakdown voltage for some of the designs considered are shown in Fig. \ref{fig:IV-noirrad} for un-irradiated sensors and in fig. \ref{fig:IV-irrad} for a fluence of $10^{15} n_{eq}/cm^2$ and compared with the standard ATLAS pixel design, with 16 guard rings and a pixel-to-trench distance of 1.1 mm. The depletion voltage has been estimated by fits to the log(C)-log(V) curve. 

The simulations indicate that the breakdown voltage exceeds by at least 100V the depletion voltage for all the configuration considered. The current is larger for 200 $\mu m$ pixel-to-trench distance with respect to the 100 $\mu m$ option, since in the first case the depleted volume extends more in the transition region. Adding one or two guard rings increases the value of breakdown voltage for un-irradiated devices, but this need is less important after irradiation, when the breakdown can occur at much higher voltage, a known effect reported in literature (see for instance \cite{weigell}).  The best performance is qualitatively obtained from a design with 2 guard rings and a 100 $\mu m$ pixel-to-trench distance.

\begin{figure}[h!] 
\centering 
\includegraphics[width=\columnwidth,height=6cm]{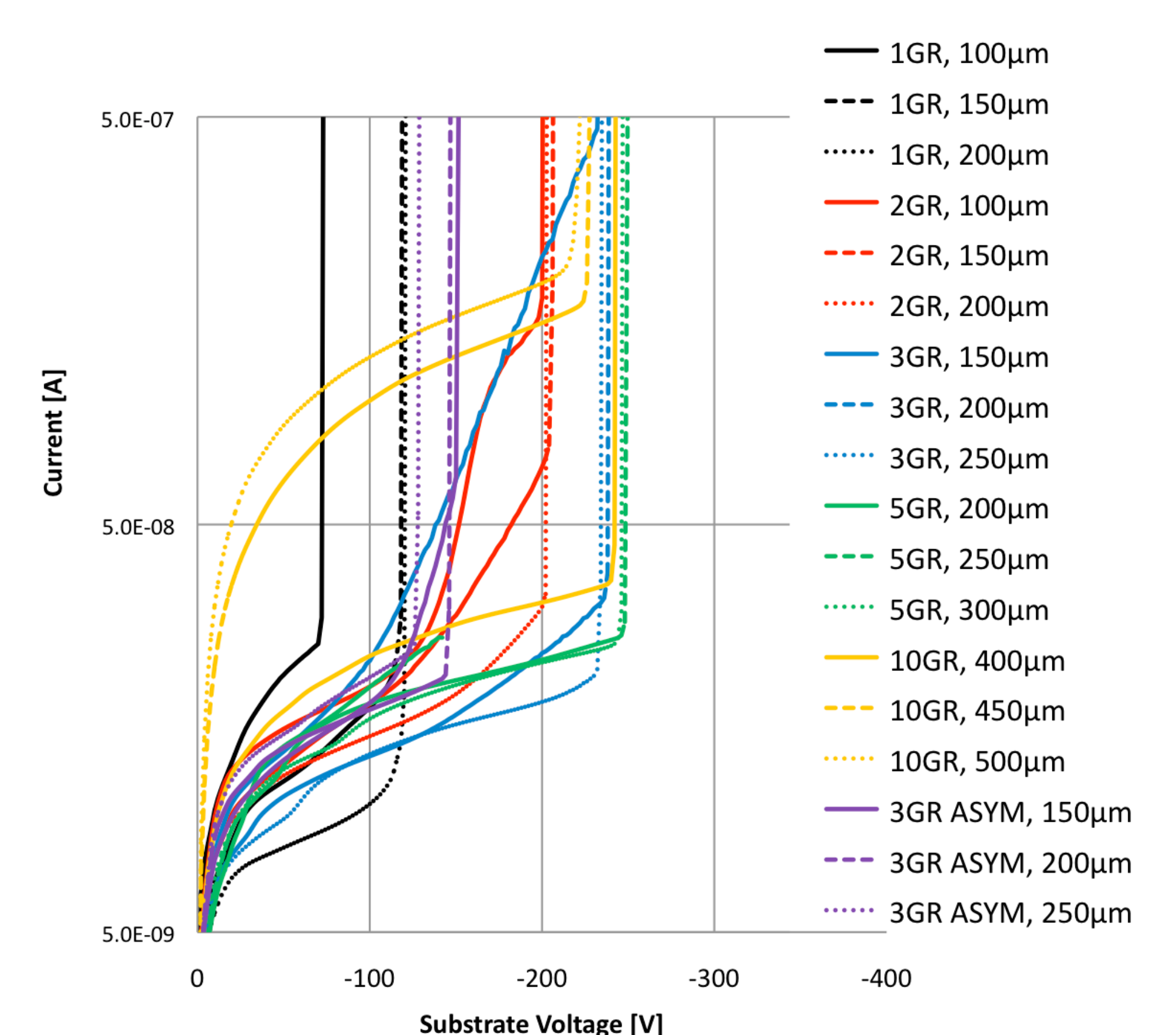}
\caption{Measured IV curves for devices with different configurations in terms of number of guard rings and pixel-to-trench distance.}
\label{fig:IV-real-devices}
\end{figure}

The measurements made on the first wafers we received are in very good agreement with the simulations. The FE-I4-design baby sensors were the first we characterized and the results in terms of current are shown in Fig. \ref{fig:IV-real-devices}, confirming a breakdown voltage well exceeding the 100V. 

\begin{figure}[h!] 
\centering 
\includegraphics[width=\columnwidth,height=6cm]{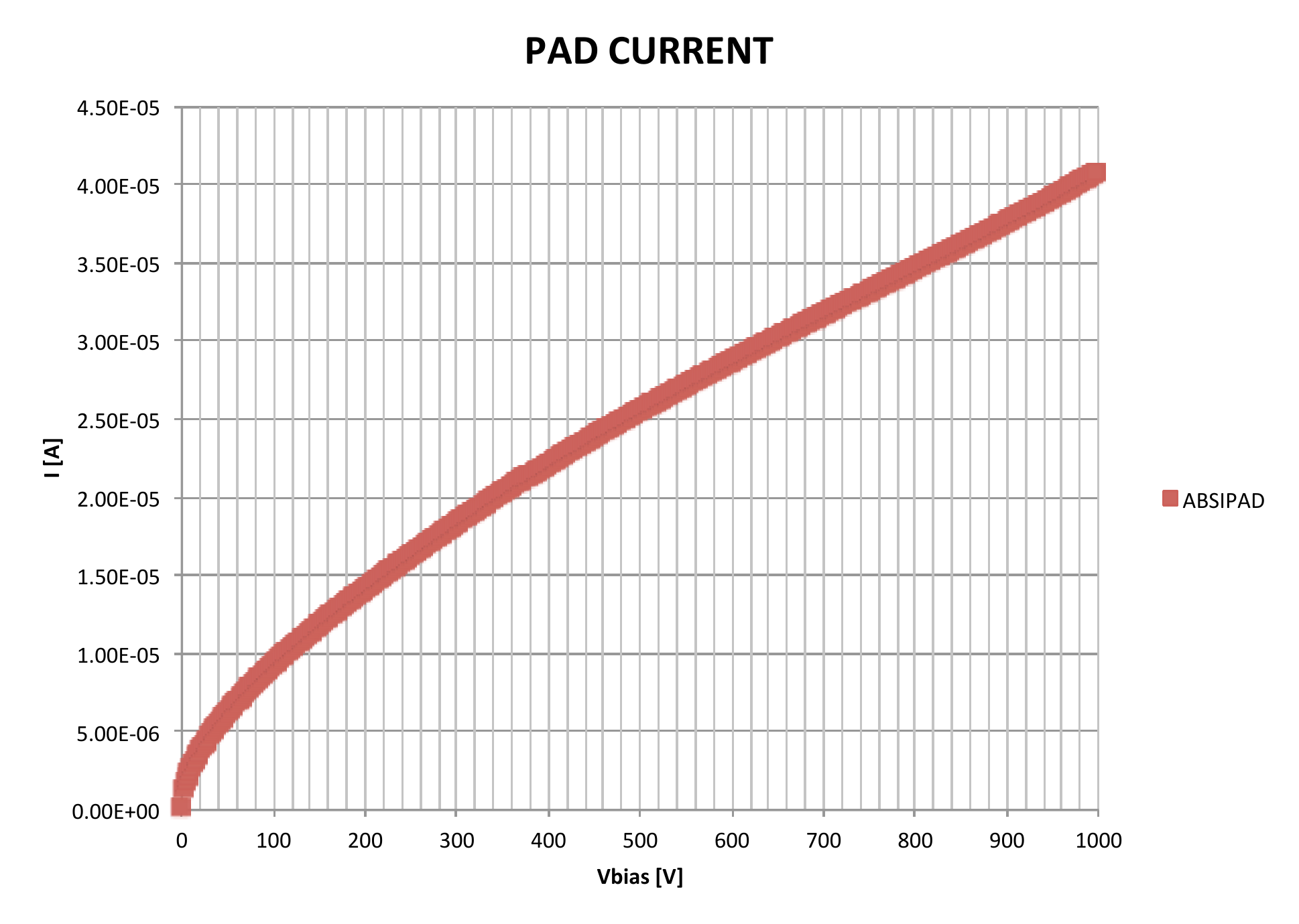}
\caption{Measured IV curves for a two-guard-rings device irradiated to a fluence of  $2.5 \times 10^{15} n_{eq}/cm^2$. The qualitative agreement with the predictions of Fig. \ref{fig:IV-irrad} is very good.}
\label{fig:IV-real-irrad}
\end{figure}

We also made preliminary irradiations with neutrons of one of the test structures featuring two guard rings to a fluence of $2.5 \times 10^{15} n_{eq}/cm^2$ in order to compare the qualitative behaviour of the leakage current and breakdown voltage with the simulations of irradiated devices. The plot, shown in Fig. \ref{fig:IV-real-irrad}, 
is in good qualitative agreement with the curves of Fig. \ref{fig:IV-irrad}.  

\begin{figure}[h!] 
\centering 
\includegraphics[width=\columnwidth,height=6cm]{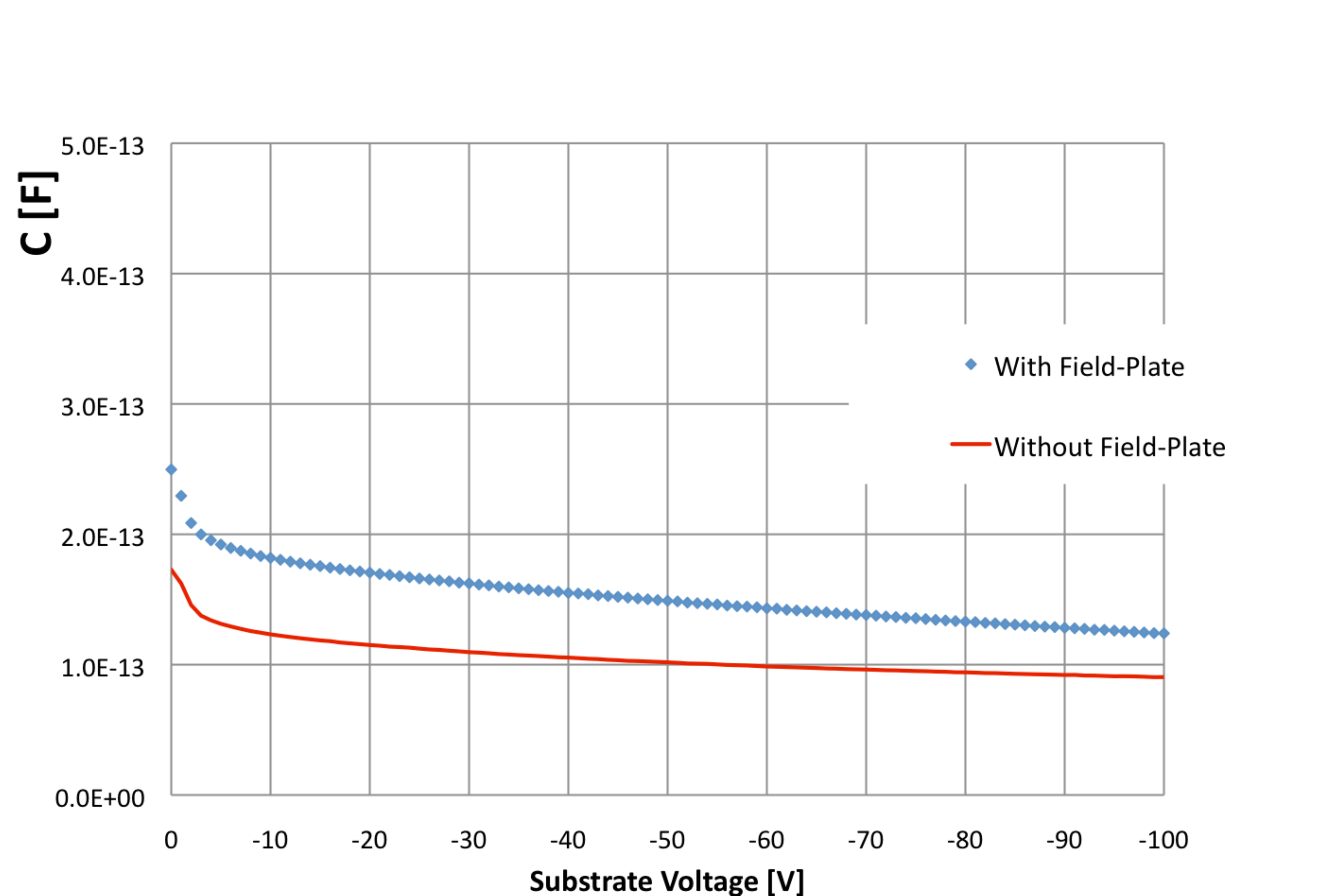}
\caption{Interpixel capacitance as a function of bias voltage measured by using a test structure with FE-I4 design. The two configurations with and without field-plate are compared.}
\label{fig:cv}
\end{figure}

The inter-pixel capacitance has also been measured with dedicated test structures which had been added to the wafer layout. In these structures a central pixel is surrounded by neighbouring pixels connected together, to make the test easier. This has been used to study the presence of field-plate effect, which can increase the inter-pixel capacitance. The  coupling is specially important due to the uniform p-spray implant. Fig. \ref{fig:cv} shows the inter-pixel capacitance as a function of bias voltage with and without field plate.

\subsection{Charge collection efficiency}
\begin{figure}[h!] 
\centering 
\includegraphics[width=\columnwidth,height=6cm]{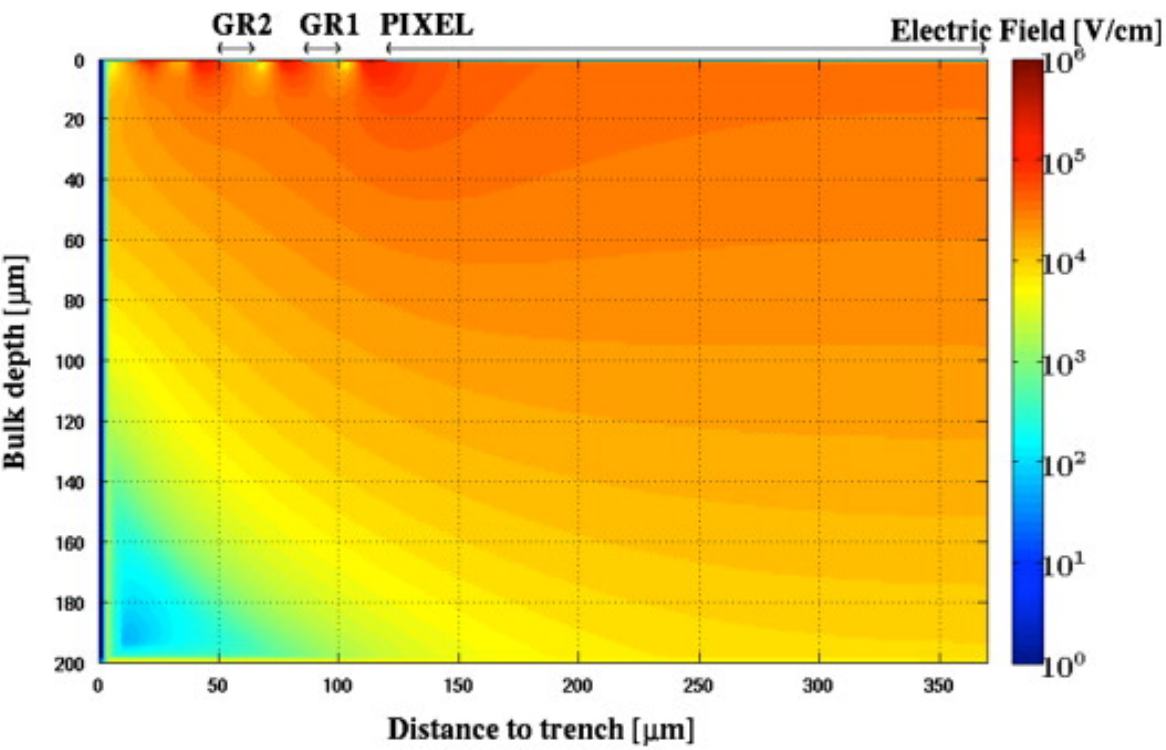}
\caption{Electric field distribution for a two guard rings device irradiated with a fluence of $10^{15}n_eq/cm^2$ and biased at 400V.}
\label{fig:field}
\end{figure}

The simulation of electric field inside the buld indicates that even after irradiation, at a bias voltage of 400V well exceeding the depletion voltage of 250V the electric field extends through the whole thickness of the sensor still reaches laterally the area below the guard rings, allowing a good charge collection in that region (see Fig. \ref{fig:field}).

\begin{figure}[h!] 
\centering 
\includegraphics[width=0.9\columnwidth,height=6cm]{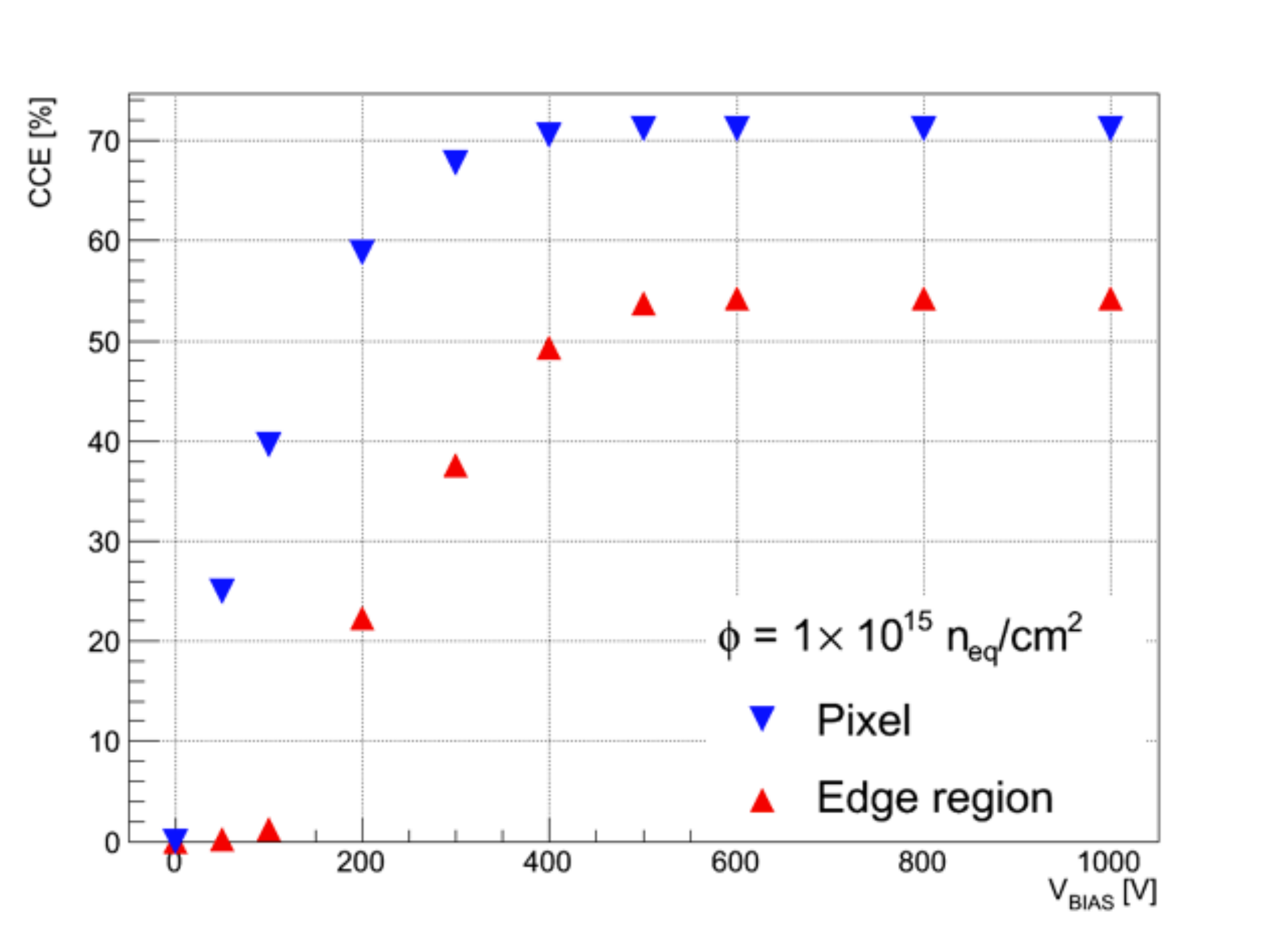}
\caption{Comparison of pixel and edge charge collection as a function of the bias voltage for a device irradiated with a fluence of $10^{15}n_eq/cm^2$. The device simulated here has no guard rings and 100 $\mu m$ distance between the pixel and the edge.}
\label{fig:cce}
\end{figure}

The charge collection efficiency (CCE) in the device has been quantitatively simulated in pre- and after-irradiation conditions, with special focus to the region between the first pixel and the edge. To generate the charge inside the bulk, a simulated 1060nm laser beam has been used, with a 2 $\mu m$-wide Gaussian spot profile. The light intensity has been tuned to match the effect of a MIP crossing the device ($\approx$ 2.6 fC) in a 10ns window, with an additional 1ns for each the ramp-up and ramp-down phases. The charge collection has been studied for the no-GR, 100 $\mu m$ trench-to-pixel distance device. Two different incident points have been considered, the first in the pixel region to be used as a reference, the second in the edge region at 50 $\mu m$ distance from the pixel. They will be referred as "pixel" and "edge" in the following. In Fig. \ref{fig:cce} the collected charge for the two incidence points is compared as a function of the bias voltage. 

At this fluence the plot indicates that more than 50\% of the photo-generated charge is collected in the edge region for a bias voltage exceeding the 400V. This is more than two-thirds of the charge collected in the pixel region, showing that the active edge principle is actually working. The maximum of the charge collection at the edge necessitates of about 500V bias voltage which is a higher value with respect to the central region, where the maximum of the efficiency is reached already at $\approx$ 400V.  The reason is that, even after the sensor thickness is fully depleted, to reach the saturation in the charge collection at the edge the electrical field still needs to extend laterally. The collection in the edge region is absent below 100V bias, since in that limit the electric field is negligible at the border.   

\section{Conclusions}
In the framework of the ATLAS PPS project in view of the LHC high-luminosity upgrade, LPNHE Paris and FBK Trento developed new n-in-p planar pixel sensors with a significantly reduced inactive region at the border. The technology, based on a deep trench running along the edge of the sensor and heavily doped to make a single structure with the back-side, has been studied with dedicated simulations before the start of the actual production and we have demonstrated that it could represent a viable option to reduce the dead-area of the devices even after a fluence comparable with that expected for the middle- and outer-layers of the tracker at the end of the HL-LHC. First preliminary measurements 
on the produced sensors show a very good agreement with simulations and look very promising. Functional tests of the devices with radioactive sources and eventually 
in beam tests will follow, after having connected a certain number of sensors to a readout electronics. 

\section{Acknowledgements}
The edgeless sensor production at FBK was supported in part by the Autonomous Province of Trento, Project MEMS2, and in part by the Italian National Institute for Nuclear Physics (INFN). Irradiation and test operations have been supported by the French CNRS/IN2P3.

\end{document}